\documentclass[12pt]{iopart}
\usepackage{epsfig}
\def\braket#1#2{\langle #1 \,\vert\, #2 \rangle}
\def\clg#1#2#3#4#5#6{\braket{#1\,#2\,;\,#3\,#4}{#5\,#6}}
\def\half{{1\over2}}

\def\ts{\textstyle}
\begin{document}
\jl{31}
\title{On the relativistic $L-S$ coupling}
\author{P Alberto, M Fiolhais and M Oliveira}
\address{Departamento de F\'{\i}sica, Universidade de Coimbra,
P-3000 Coimbra, Portugal}

\begin{abstract} The fact that the Dirac equation is linear in the space
and time derivatives leads to the coupling of spin and orbital angular
momenta that is of a pure relativistic nature.  We illustrate this fact
by computing the solutions of the Dirac equation in an infinite
spherical well, which allows to go from the relativistic to the
non-relativistic limit by just varying the radius of the well.
\end{abstract}

\submitted

\section{Introduction}

The effect of spin-orbit coupling is well known from elementary quantum
mechanics and atomic physics:  it arises from the interaction between
the spin of an electron in an atom and the magnetic field created by the
orbiting nucleus in the rest frame of the electron.  This magnetic field
is related to the electric field created by the nucleus in its rest
frame. If this field is a spherical electrostatic field, the
interaction hamiltonian is given by
\begin{equation}
\label{spin-orbit_energy} H_{\rm{spin-orbit}}=
{e\over 2m^2c^2}{1\over r}{{\rm d} V\over{\rm d} r}\vec S\cdot\vec L\ .
\end{equation}
Here, as usual, $\vec S$ and $\vec L$ are the spin and orbital momentum
operators for the electron, $m$ and $e$ stand for the electron charge
and mass, $c$ is the speed of light in the vacuum and $V(r)$ is the
electrostatic potential of the atomic nucleus.  For one-electron atom,
the formula \eref{spin-orbit_energy} is exact, otherwise $V(r)$ can be
thought as an approximation to an average radial potential experienced
by the electron.  Equation \eref{spin-orbit_energy} is obtained in the
non-relativistic limit (electron velocity is small compared to $c$ ---
see, for instance, \cite{Bethe_Jackiw}) and so it is used in the
non-relativistic description of an electron, i.e., by adding it to the
Hamiltonian in the Schr\"odinger equation.

In this paper we propose to examine a similar coupling that arises due
to the relativistic treatment of the electron (i.e. using the Dirac
equation) {\it even in the absence of an external field}.  This is a
consequence of the linearity of the Dirac equation in the space
derivatives (and thus in the linear momentum operator $\vec p$) and from
the related fact that one needs a 4-component spinor to describe the
electron.  We will make the relativistic nature of this coupling
apparent by solving the Dirac equation in an infinite spherical
potential well.  Although the particle motion inside the well is free,
the relativistic $L-S$ coupling exists and vanishes only in the
non-relativistic limit, which we are able to approach continuously by
varying the well radius.  In this limit the two-component spinor
description is valid.

A comparison between relativistic and non-relativistic solutions was
already studied in \cite{pedro_parbox} for a one-dimensional infinite
square well potential.  In the present paper we use the same
procedures as in
\cite{pedro_parbox} to provide a bridge between known relativistic and
non-relativistic solutions in the 3-dimensional spherical case, with
special emphasis on the $L-S$ coupling.  Berry and Mondragon
\cite{Berry} have also applied similar methods in the framework of the
Dirac equation in two spatial dimensions.

In section 2 we pedagogically review the solutions of the free Dirac
equation with spherical symmetry, in a slightly different fashion from
the usual treatments, emphasizing the role of the $L-S$ coupling term
and its consequences for the set of quantum numbers of the solution.  In
section 3 we solve the Dirac equation for a spherical potential well and
compare it to the non-relativistic solution of the corresponding
Schr\"odinger equation for several well radii. Technical details,
included for completeness, are mostly left to Appendices.

\section{Solutions of the free Dirac equation with spherical symmetry}

The free Dirac equation for a spin-$\case12$ particle with mass $m$ is a
matrix equation for 4-component spinors $\Psi$ given by
\begin{equation}
\label{Dirac_eq}
\rmi\,\hbar{\partial\Psi\over\partial t}=
\vec\alpha\cdot\vec p\, c\,\Psi+\beta mc^2\,\Psi
\end{equation}
where
$\vec p=-\rmi\,\hbar\vec\nabla$ is the linear momentum operator, and
$\vec\alpha$ and $\beta$, in the usual representation, are the $4\times
4$ matrices
\begin{equation}
\vec\alpha=\pmatrix{0&\vec\sigma\cr
\vec\sigma&0\cr}\qquad\beta=\pmatrix{I&0\cr 0&-I\cr}\ .
\end{equation}
Here $I$ is the $2\times 2$ unit matrix and $\vec\sigma$ denotes the
three Pauli matrices $\sigma_i\quad i=1,2,3$ obeying the relations
\begin{equation}
\label{sig_i_sig_j}
\sigma_i\sigma_j=\delta_{ij}+\rmi\,\varepsilon_{ijk}\sigma_k\qquad
i,j=1,2,3
\end{equation}
where $\varepsilon_{ijk}$ is the anti-symmetric Levi-Civita tensor
($\varepsilon_{123}=1$) and summation over repeated indexes is implied.

Using \eref{sig_i_sig_j} we can obtain the following general property of
the $\alpha$ matrices
\begin{equation}
\label{alpha_A_alpha_B}
\vec\alpha\cdot\vec A\,\,\vec\alpha\cdot\vec B=\vec A\cdot\vec
B+\rmi\,\vec A\times\vec B\cdot\vec\Sigma\ ,
\end{equation}
where $\vec A$ and $\vec B$ are two arbitrary vectors whose components
commute with the matrices $\alpha_i$ and
\[
\vec\Sigma=\pmatrix{\vec\sigma&0\cr 0&\vec\sigma\cr}
\]
is the 4-dimensional analog of the Pauli matrices.  Using
\eref{alpha_A_alpha_B} and $\vec\alpha\cdot\hat r\,\vec\alpha\cdot\hat
r=I$ (here, of course, $I$ stands for the $4\times 4$ unit matrix), one
can write
\begin{eqnarray}
\label{alpha_p} \vec\alpha\cdot\vec p&=
\vec\alpha\cdot\hat r\,\,\vec\alpha\cdot\hat r\,\,\vec\alpha\cdot
\vec p\nonumber\\
&=\vec\alpha\cdot\hat r\,(\hat r\cdot\vec p+\rmi\,\hat r\times\vec p
\cdot\vec\Sigma)\nonumber\\
&=\vec\alpha\cdot\hat r\,(\hat
r\cdot\vec p+{\rmi\,\over r}\vec L\cdot\vec\Sigma)\,,
\end{eqnarray}
where $r=|\vec r|$, $\hat r=\vec r/r$ and $\vec L=\vec r\times\vec p$ is
the orbital angular momentum operator.  Inserting \eref{alpha_p} into
the Dirac equation \eref{Dirac_eq} we get
\begin{equation}
\label{Dirac_eqLS}
\rmi\,\hbar{\partial\Psi\over\partial t}=
\vec\alpha\cdot\hat r\,(\hat r\cdot\vec p+
{\rmi\,\over r}\vec L\cdot\vec\Sigma)\, c\,\Psi+\beta mc^2\,\Psi\ .
\end{equation}
Since the spin angular momentum operator in the Dirac formalism is $\vec
S={\hbar\over2}\,\vec\Sigma$ the last expression contains a term
involving the dot product $\vec L\cdot\vec S$ as in the spin-orbit term
\eref{spin-orbit_energy}.  This term is responsible for the $L-S$
coupling\footnote{We prefer to use this name because there is no orbital
motion for a free particle} for a relativistic spin-$\case12$ particle
even in the absence of an external potential.  Clearly this fact is
connected to the spinor structure of the wave function for
spin-$\case12$ particles and to the linearity of the Dirac equation,
which leads to the appearance of the term $\vec\alpha\cdot\vec p$. In
the Klein-Gordon equation, which is quadratic in the space derivatives,
there is no such coupling.

We can gain further insight into the origin of this effect if we write
the spinor $\Psi$ in \eref{Dirac_eqLS} as a set of two-component spinors
$\chi'$ and $\varphi'$:
\begin{equation}
\Psi=\pmatrix{\chi'\cr
              \varphi'\cr}\ .
\end{equation}
From the block off-diagonal form of the $\alpha_i$ matrices one sees
that the term $\vec\alpha\cdot\vec p$ mixes the spinors $\chi'$ and
$\varphi'$.  In the literature (see, for instance, Bjorken and Drell
\cite{BjorkenDrell}) this kind of term is referred to as ``odd" as
opposed to the terms like $\beta mc^2$ which are called ``even", since
they don't mix upper and lower two-component spinors.  This distinction
is important when one goes to the non-relativistic limit of the Dirac
equation using the Foldy-Wouthuysen transformation
\cite{BjorkenDrell,FoldyWouthysen} which aims at eliminating the odd
terms through a unitary transformation and so decouple the upper and
lower spinors.  In this way one can regain the (Pauli) non-relativistic
description of a spin-$\case12$ particle.  The fact that the term
responsible for the $L-S$ coupling is odd indicates that this
relativistic effect is related to the four-component spinor structure,
i.e., to the existence of two non-zero spinor components $\chi'$ and
$\varphi'$ of the wave function.  This will be shown in the following.

Let us consider the stationary solutions of the Dirac equation by
writing the spinor $\Psi$ in the form
\begin{equation}
\Psi=e^{-\rmi\,E/\hbar\,t}\pmatrix{\chi\cr
              \varphi\cr}\ ,
\end{equation}
where $E$ is the total (kinetic plus rest) energy of the fermion.
Inserting this expression into \eref{Dirac_eqLS} we get two equations
for the spinors $\varphi$ and $\chi$
\begin{eqnarray}
\label{chi_equation}
(E-mc^2)\chi= \vec\sigma\cdot\hat r\,(\hat r\cdot\vec p+
{\rmi\,\over r}\vec L\cdot\vec\sigma)\, c\,\varphi\\
\label{phi_equation}
(E+mc^2)\varphi= \vec\sigma\cdot\hat r\,(\hat r\cdot\vec p+
{\rmi\,\over r}\vec L\cdot\vec\sigma)\, c\,\chi.
\end{eqnarray}
From these equations the $L-S$ coupling of the spinors $\varphi$ and
$\chi$ becomes apparent. Let us consider solutions with spherical
symmetry of these equations. In Appendix A a derivation slightly
different from the one used in most textbooks is presented.  It is
shown that the spinors can be written as products of a radial and an
angular function as
\begin{eqnarray}
\chi=\rmi\,G_{j\ell}(r)\Phi_{j\ell m}(\theta,\phi)\ ,\\
\label{ansatz_phi_2}
\varphi=-F_{j\ell'}(r)\Phi_{j\ell' m}(\theta,\phi)=F_{j\ell'}(r)
\vec\sigma\cdot\hat r\,\Phi_{j\ell m}\ ,
\end{eqnarray}
where $\ell'=\ell-\kappa/|\kappa|$, $\kappa$ being a non-zero quantum
number which has a different sign according to the way the spin couples
to the orbital angular momentum (see \eref{def_kappa}).  Since
$\ell'\not=\ell$ the whole spinor $\Psi$ is not an eigenstate of the
orbital angular momentum operator $\vec L^2$.  The good quantum numbers
are $j$ (total angular momentum quantum number), $s=\case12$, $m$ (see
Appendix A) and parity.  This is due to the $L-S$ term mentioned above.

It is interesting to look at the non-relativistic limit of the
equation \eref{phi_equation}. If we divide it by $mc^2$, we obtain
\begin{equation}
\label{eq_phi2}
\bigg({E\over mc^2}+1\bigg)\varphi=
{1\over mc}\vec\sigma\cdot\vec p\chi,
\end{equation}
using the fact that $\vec\sigma\cdot\hat r(\hat r\cdot\vec p+
{\rmi\,\over r}\vec L\cdot\vec\sigma)\, c=\vec\sigma\cdot\vec p\,c$.  In
the non-relativistic limit, the linear momenta of the dominant
plane-wave components of $\chi$ (obtained through a Fourier
decomposition) are much smaller than $mc$, which implies, from
\eref{eq_phi2}, that $\varphi$ disappears in that limit.  Since the
angular part of $\varphi$ contains only geometrical information, we can
conclude (see equation \eref{ansatz_phi_2}) that $F_{j\ell'}$ vanishes
in the non-relativistic limit and one recovers the two-component spinor
description of a spin-$\half$ particle.

Interestingly enough, in the ultra-relativistic limit, where $E+mc^2\sim
E-mc^2\sim E$, we can again recover the two-component description, since
in this case we can choose the spinors $\chi$ and $\varphi$ to be
eigenstates of the helicity operator
\[
\vec\sigma\cdot{\vec p\,c\over E}=\vec\sigma\cdot{\hat p}\,,
\]
where $\hat p=\vec p/|\vec p|$, with eigenvalues $\pm 1$, as can be seen
from equations \eref{chi_equation} and \eref{phi_equation}.  This
implies that $\chi=\pm\varphi$ and therefore we may construct two
two-component spinors for each value of the helicity (see, for
instance, Itzykson and Zuber \cite{ItzyksonZuber}).

If we differentiate once the coupled first order differential equations
for $G_{j\ell}$ and $F_{j\ell'}$ derived in Appendix A (equations
\eref{eq_G(r)} and \eref{eq_F(r)}), one gets
\begin{eqnarray}
\label{eq_G(r)_2}
{{\rm d}^2 G_{j\ell}\over {\rm d}r^2}+{2\over r}\,{{\rm d} G_{j\ell}
\over {\rm d}r}-{\ell(\ell+1)\over r^2}G_{j\ell}+{E^2-m^2c^4\over
(\hbar c)^2}G_{j\ell}=0\\
\label{eq_F(r)_2}
{{\rm d}^2 F_{j\ell'}\over {\rm d}r^2}+{2\over r}\,{{\rm d} F_{j\ell'}
\over {\rm d}r}-{\ell'(\ell'+1)\over r^2}F_{j\ell'}+{E^2-m^2c^4\over
(\hbar c)^2}F_{j\ell'}=0 .
\end{eqnarray}
Notice that, although in each equation only $\ell$ or $\ell'$ appear
explicitly, the radial functions depend also on $j$ through the energy
$E$.  These are the differential equations which have to be solved in
order to get the radial functions $G_{j\ell}$ and $F_{j\ell'}$.

In the non-relativistic limit, since $F_{j\ell'}$ vanishes, $\ell$ is
again a good quantum number, i.e., the $L-S$ coupling disappears.
Moreover, since in this case $G_{j\ell}$ only depends on $\ell$, we can
construct the standard non-relativistic solution taking the linear
combination

\begin{equation}
G_{\ell}\sum_{j\,m}\clg\ell{m_\ell}\half{m_s}jm\Phi_{j\ell m}=
G_{\ell}Y_{\ell m_\ell}\chi_{m_s}\ ,
\end{equation}
where we dropped the index $j$ in the radial function and used the
definition \eref{phi_jlm} of $\Phi_{j\ell m}$ and an orthogonality
property of the Clebsch-Gordan coefficients.

The differential equations \eref{eq_G(r)} and \eref{eq_F(r)} can
be extended to include interactions with spherical external potentials
$V(r)$ and $m(r)$, which are respectively a time component of a
four-vector (affecting the energy) and a Lorentz scalar (affecting the
mass).  This is done by the replacements $E\longrightarrow E-V(r)$ and
$m\longrightarrow m(r)$.

\section{Solution of the Dirac equation in an infinite spherical well}

In order to show numerically the relativistic nature of the $L-S$
coupling described in the preceding section, we are going now to compute
the positive energy solutions of the Dirac equation for an infinite
spherical well. As we will show, we can go, in a natural
way, from a relativistic to a non-relativistic situation by changing the
radius of the potential.  The boundary conditions at the wall of the
potential provide a discrete energy spectrum which allows a clear
picture of the non-relativistic limit.

To solve the Dirac equation in such a potential, one has to avoid any
complications due to the negative energy states when trying to localize
a spin-$\half$ particle within a distance of the order of its Compton
wavelength $\hbar/(mc)$ or less (this is the case for
confined relativistic particles), one example of which is the Klein
paradox (see, for instance, \cite{BjorkenDrell}).  In other words, we
want to retain the r\^ole of the Dirac equation as a one-particle
equation  in the presence of a infinite external potential.
This is accomplished by defining a Lorentz scalar potential, i.e., a
mass-like potential, having the form
\begin{equation}
\label{def_m(r)}
m(r)=\cases{m&$r<R$\cr
            \infty & $r>R$\cr}\ ,
\end{equation}
where $m$ is the mass of the particle.  The effect of this potential is
to prevent the particle from propagating outside the well, meaning that
its wave function is identically zero there.  Inside, it behaves as a
free particle of mass $m$.  A potential like \eref{def_m(r)}, usually
with $m=0$, has been used to describe confined quarks as constituents of
the nucleon in the MIT bag model (see, e.g., \cite{Thomas} for a review
of this and related models).

The boundary condition for the wave function at the boundary ($r=R$)
cannot be obtained by requiring its continuity, since, being the Dirac
equation a first-order differential equation, the potential
\eref{def_m(r)} implies that there is an infinite jump in the derivative
of $\Psi$ (i.e., in the radial derivatives of $G$ and $F$)
when the boundary of the well is crossed.  This jump obviously would not
exist if $\Psi$ were continuous.  Another and most natural alternative
is to demand that the probability current flux at the boundary is zero.
As it is shown in \cite{Berry}, this is also a necessary
condition to assure the hermiticity of the kinetic part of the Dirac
hamiltonian within the well.  This can be achieved by the condition
\begin{equation}
\label{bound_cond}
-\rmi\,\beta\vec\alpha\cdot\hat r\Psi=\Psi \quad {\rm at}\quad r=R .
\end{equation}
In fact, if one multiplies this equation on the left by
$\Psi^\dagger\beta$
and its hermitian conjugate on the right by $\beta\Psi$ one gets
$-\rmi\,\Psi^\dagger\vec\alpha\cdot\hat r\Psi=\Psi^\dagger\beta\Psi$ and
$-\rmi\,\Psi^\dagger\vec\alpha\cdot\hat r\Psi=
\Psi^\dagger\beta\Psi$ at $r=R$.
These two equations imply that $\Psi^\dagger\beta\Psi$ and
$\Psi^\dagger\vec\alpha\cdot\hat r\Psi$ are zero at $r=R$.

The expression $\Psi^\dagger\vec\alpha\cdot\hat r\Psi$ can also be
written as $\vec j\cdot\hat r/c$, where $\vec
j=\Psi^\dagger\vec\alpha\Psi c$ is the probability current density for
the particle described by the wave function $\Psi$.  Instead of the
current flux we can look at the value of $\Psi^\dagger\beta\Psi$ at the
boundary:  indeed, since $\Psi$ is zero for $r>R$, we may as well
summarize the effect of the boundary condition \eref{bound_cond} by
saying that $\Psi^\dagger\beta\Psi$ is continuous for any value of $r$.

Having established the boundary condition, we proceed now to compute the
radial functions.  This is done in Appendix C. The full spinor $\Psi$
reads
\begin{equation}
\label{Psi_complete}
\fl
\Psi_{j\kappa m}(r,\theta,\phi,t)=A\,e^{-\rmi\,E/\hbar\,t}
\pmatrix{\rmi\,j_\ell({\sqrt{E^2-m^2c^4}\over\hbar c}\,r)
\Phi_{j\ell m}(\theta,\phi)  \cr
-{\kappa\over|\kappa|}\sqrt{E-mc^2\over E+mc^2}\,\,j_{\ell'}
({\sqrt{E^2-m^2c^4}\over\hbar c}\,r)\Phi_{j\ell' m}(\theta,\phi) \cr}\ ,
\end{equation}
where $A$ is determined from normalization.  In order to obtain the
energy spectrum, we apply the boundary condition \eref{bound_cond} to
the spinor \eref{Psi_complete}.  This gives rise to an equation relating
the two radial functions (see Appendix C)
\begin{equation}
j_\ell(X)=-{\kappa\over|\kappa|}\sqrt{E-mc^2\over E+mc^2}\,\,
j_{\ell'}(X)
\end{equation}
where $X={\sqrt{E^2-m^2c^4}/(\hbar c)}\,R$. It can be
written, in a more convenient way, in terms of the scaled quantities
$y=(E-mc^2)/(mc^2)$ and $x_R=R/L_0$, with $L_0=\hbar/(m c)$.  These are
the kinetic energy in units of $m c^2$ and the well radius in units of
the Compton wavelength, respectively.  We get then
\begin{equation}
\label{j_l_j_l'_eq}
j_\ell(x_R\sqrt{y^2+2 y})=-{\kappa\over|\kappa|}\sqrt{y\over y+2}\,\,
j_{\ell'}(x_R\sqrt{y^2+2 y}) \ .
\end{equation}
 This equation is solved numerically for $y$ as a function of $x_R$ for
a given set of $\ell$, $\ell'$ and $\kappa$.
\begin{figure}[hbt]
\begin{center}
\epsfclipon
\epsfig{width=\textwidth,file=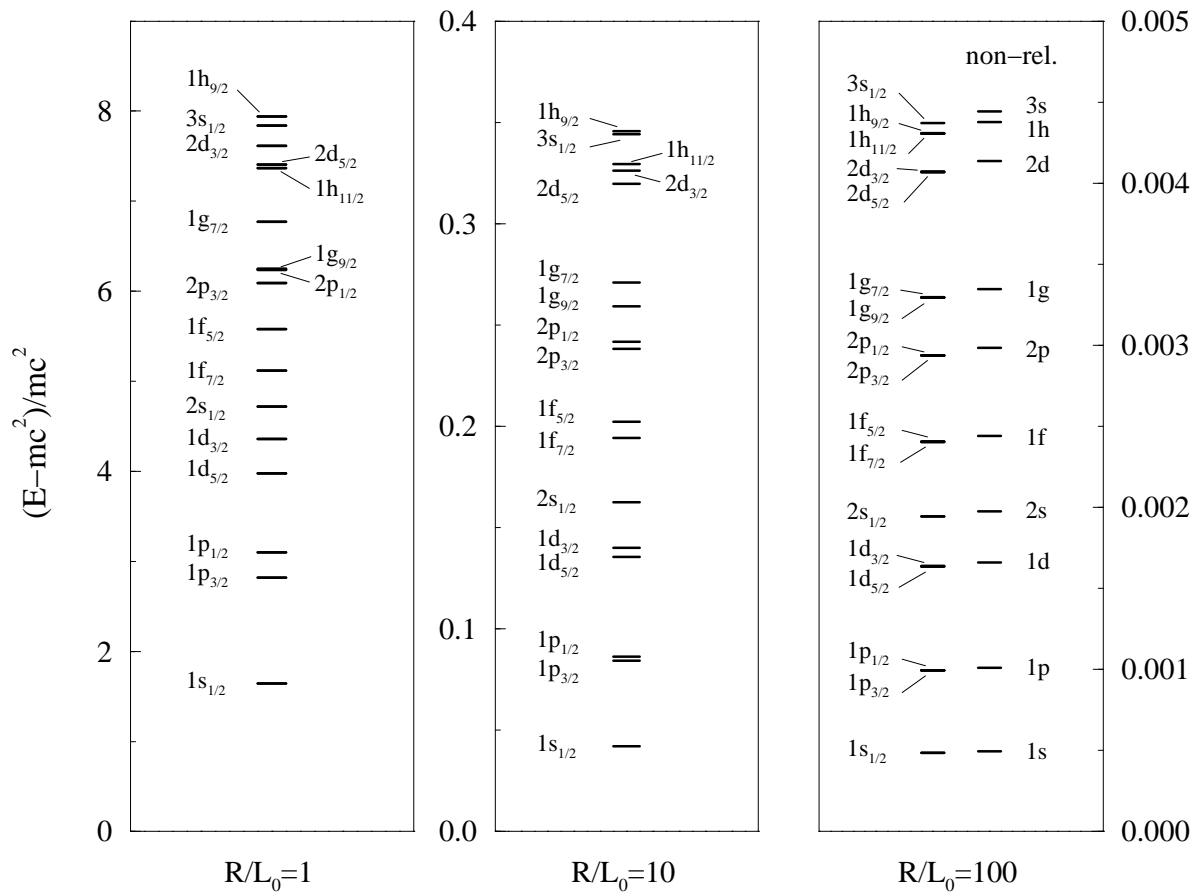}
\end{center}
\caption{The first 17 scaled kinetic energy levels obtained by solving
equation \eref{j_l_j_l'_eq} for values of $x_R$ equal to 1, 10 and 100.
In the last plot the first 10 scaled kinetic energy levels obtained by
solving the Schr\"odinger equation for an infinite spherical potential
well of radius $R=100L_0$ are also shown.}
\end{figure}
The results are presented in Figure 1. We plot the first values of $y$
up to $\ell=5$ for three values of $x_R$.  The energy levels are labeled
in standard spectroscopic notation $n\,\ell_j$, where $n$ denotes the
$n$th solution for a given set of $\ell$ and $j$.  For $x_R=100$ the
non-relativistic results, using the notation $n\,\ell$, are also
presented.  The non-relativistic spectrum is obtained by solving the
Schr\"odinger equation for a particle of mass $m$ in an infinite
spherical potential well of radius $R=100L_0$.  The solutions can be
found, for instance, in the quantum mechanics textbook of Landau
\cite{Landau} (in this case, there is no spin-orbit coupling of the type
\eref{spin-orbit_energy} because the potential is zero inside the
well).
The radial functions are spherical Bessel functions subject to the
boundary condition
\begin{equation}
\label{non-rel_condition}
j_\ell(kR)=0\ ,\quad k={\sqrt{2mE_k}\over\hbar},
\end{equation}
where $E_k$ is the kinetic energy of the particle.  Notice that, in the
non-relativistic limit, $y\ll 1$, equation \eref{j_l_j_l'_eq} reduces to
\eref{non-rel_condition} since
\begin{equation}
x_R\sqrt{y^2+2 y}\sim x_R\sqrt{2 y}={\sqrt{2mE_k}\over\hbar}\,R\ ,
\end{equation}
and the factor $\sqrt{y/(y+2)}$ goes to zero in this limit.

Analyzing Figure 1, we see that, as the radius of the well increases,
the energy levels with the same $\ell$ start grouping until they become
degenerate and almost identical to the corresponding non-relativistic
values.  This effect is more pronounced for the states with higher
$\ell$ (notice the behaviour of the $1h_{9/2}$ and $1h_{11/2}$ states).
So we can conclude that going from a radius $R=L_0$ to a radius
$R=100L_0$ the $L-S$ coupling effect fades away and $j$ is no longer
needed to classify the eigenstates of the system, and instead the
orbital momentum quantum number $\ell$ emerges as the relevant quantum
number.  Since the boundary condition \eref{j_l_j_l'_eq} effectively
imposes a (maximum) value for the wavelength of the wave function and
thereby a (minimum) value for the energy through the De Broglie
relation, increasing the radius of the well amounts to decreasing the
energy until we reach non-relativistic values for $R=100L_0$.  Notice
that for the higher levels, for this value of $R$, even though there is
not a perfect match with the non-relativistic energy values, the
vanishing of the $L-S$ coupling is a fact.  The crucial scale here is
the Compton wavelength $L_0=\hbar/(m c)$, determining the relativistic
nature of the solution through the well radius.

\vskip2mm
In summary, we have showed numerically the relativistic nature
of the $L-S$ coupling in the Dirac equation by computing its solutions
for a particle with mass $m$ in an infinite spherical potential well of
radius $R$ and making $R$ sufficiently big as to produce
non-relativistic solutions.

\ack

This work was supported by the Project PRAXIS/PCEX/C/FIS/6/96.

\appendix

\section{}

In this Appendix the radial equations for the Dirac equation are
derived.  We first write $\varphi$ and $\chi$ in equations
\eref{chi_equation} and \eref{phi_equation} as products of a radial
function and a function of the angular coordinates $\theta$ and $\phi$.
To be able to get ordinary differential equations for the radial
functions, the angular function must be an eigenstate of the operator
$\vec L\cdot\vec\sigma=(\vec J^2-\vec L^2-\vec S^2)/\hbar$ (where $\vec
J=\vec L+\vec S$ denotes the total angular momentum), which acts only on
the angular coordinates.  Accordingly, the angular function,
$\Phi_{j\ell m}$, reads
\begin{equation}
\label{phi_jlm}
\Phi_{j\ell m}(\theta,\phi)=
\sum_{m_\ell=-\ell}^\ell\sum_{m_s=-\half}^\half
\clg\ell{m_\ell}\half{m_s}jm Y_{\ell m_\ell}(\theta,\phi)\chi_{m_s}
\end{equation}
where $Y_{\ell m}(\theta,\phi)$ is the spherical harmonic with
quantum numbers $\ell$ and $m$, $\chi_{m_s}$ the two-component spinors
\[
\chi_{\half}=\pmatrix{1\cr 0\cr} \qquad \chi_{-\half}=
\pmatrix{0\cr 1\cr}\ ,
\]
and $\clg\ell{m_\ell}\half{m_s}jm$ is a Clebsch-Gordan
coefficient.  The wave function \eref{phi_jlm} is an eigenstate of $\vec
J^2$, $\vec L^2$, $\vec S^2$ and $J_z$ with eigenvalues $\hbar^2j(j+1)$,
$\hbar^2\ell(\ell+1)$, $\hbar^2\half(\half+1)=\hbar^2{3\over4}$ and
$\hbar m$ respectively.  Therefore we have
\begin{eqnarray}
\vec L\cdot\vec\sigma\,\Phi_{j\ell m}&=\hbar[j(j+1)+
\ell(\ell+1)-{3/4}]\Phi_{j\ell m}=\nonumber\\
&=-\hbar(1+\kappa)\Phi_{j\ell m}
\end{eqnarray}
with
\begin{equation}
\label{def_kappa}
\kappa=\left\{\matrix{\hfill-(\ell+1)=-(j+\half)&\qquad j=\ell+\half\cr
              \hfill\ell=\phantom{-(}j+\half\phantom{)}&
              \qquad j=\ell-\half\cr}\right.\ .
\end{equation}
For a fixed $j$, the quantum number $\kappa$ takes into account
the two different possibilities for $\ell$, namely $\ell=j\pm\half$, by
just changing its sign. It also satisfies the equality
$\kappa(\kappa+1)=\ell(\ell+1)$ for a certain $\ell$.  Thus $\kappa$ can
be considered as an alternative quantum number for the wave function
$\Phi_{j\ell m}$ replacing $\ell$.  The corresponding operator is
$-(\hbar+\vec L\cdot\vec\sigma)$.  Note that for $\ell=0$ only one value
of $\kappa$ is defined (-1). Wave functions with a fixed $j$ but
different $\ell$'s have opposite parity, since $\ell=j\pm\half$
and parity is given by $(-1)^\ell$). Using standard
notation, $\varphi$ and $\chi$ are then written as
\begin{eqnarray}
\label{ansatz_chi}
\chi=\rmi\,G_{j\ell}(r)\Phi_{j\ell m}(\theta,\phi)\ ,\\
\label{ansatz_phi}
\varphi=-F_{j\ell'}(r)\Phi_{j\ell' m}(\theta,\phi)\ .
\end{eqnarray}
The quantum number $\ell'$
of the lower component $\varphi$ can be found by applying the operator
$\vec\sigma\cdot\hat r(\hat r\cdot\vec p+ {\rmi\,\over r}\vec
L\cdot\vec\sigma) c$ to $\chi$ (see equation
\eref{phi_equation}), giving
\begin{eqnarray}
\vec\sigma\cdot\hat r\,(\hat r\cdot\vec p+
{\rmi\,\over r}\vec L\cdot\vec\sigma)\, c\,\chi&=
c\,\vec\sigma\cdot\hat r(-\rmi\,\hbar{\partial\hfil\over\partial r}+
{\rmi\,\over r}\vec L\cdot\vec\sigma) \rmi\,G_{j\ell}(r)\,\Phi_{j\ell m}
\nonumber\\
\label{phi_r}
&=\hbar c\bigg[{{\rm d} G_{j\ell}\over{\rm d}r}+(1+\kappa) {G_{j\ell}
\over r}\bigg]\vec\sigma\cdot\hat r\,\Phi_{j\ell m}\ .
\end{eqnarray}
The effect of $\vec\sigma\cdot\hat r$ over $\Phi_{j\ell m}$ can be
computed using the tensor properties of $\vec\sigma$ and $\hat r$
(see Appendix B), yielding
\begin{equation}
\label{sigma_r_phi_jlm}
\vec\sigma\cdot\hat r\,\Phi_{j\ell m}=-\Phi_{j\ell'm}\ ,
\end{equation}
where $\ell'$ is given by
\begin{equation}
\ell'=\cases{\ell+1&if $\quad j=\ell+\half$\cr
             \ell-1&if $\quad j=\ell-\half$.\cr}
\end{equation}
Note that $\ell'$ is related to $\ell$ by $\ell'=\ell-\kappa/|\kappa|$.
 If we define the operator
\begin{equation}
K=\pmatrix{-(\hbar+\vec L\cdot\vec\sigma)&0\cr
0&\hbar+\vec L\cdot\vec\sigma\cr}\ ,
\end{equation}
$\Psi$ will be an eigenstate of $K$ with eigenvalue $\kappa$.  Thus
$\kappa$ is also a good quantum number.  From \eref{phi_equation} and
\eref{phi_r} we can write $\varphi$ in \eref{ansatz_phi} in the form
\begin{equation}
\varphi=F_{j\ell'}(r)\,\vec\sigma\cdot\hat r\,\Phi_{j\ell m}
(\theta,\phi)\ .
\end{equation}

The radial functions $G_{j\ell}(r)$ and $F_{j\ell'}(r)$ satisfy the
coupled differential equations (see equations \eref{chi_equation} and
\eref{phi_equation})
\begin{eqnarray}
\label{eq_G(r)}
(E-mc^2)G_{j\ell}= -\hbar c\bigg[{{\rm d} F_{j\ell'}\over{\rm d}r}+
(1+\kappa') {F_{j\ell'}
\over r}\bigg]\\
\label{eq_F(r)}
(E+mc^2)F_{j\ell'}= \hbar c\bigg[{{\rm d} G_{j\ell}\over{\rm d}r}+
(1+\kappa) {G_{j\ell}\over r}\bigg]\ ,
\end{eqnarray}
where $\kappa'$ is related to $\ell'$ in the same way as in
\eref{def_kappa} (giving the relation $\kappa'=-\kappa$) and the
relation $\vec\sigma\cdot\hat r\Phi_{j\ell'm}=-\Phi_{j\ell m}$ was used
(note that $(\vec\sigma\cdot\hat r)^2=I$).


\section{}

In this Appendix we will derive expression \eref{sigma_r_phi_jlm} by
calculation the matrix element
\begin{equation}
\Phi_{j'\ell' m'}^\dagger\,
\vec\sigma\cdot\hat r\,\Phi_{j\ell m}\ ,
\end{equation}
where `$\dagger$' stands for hermitian conjugate.  Since both
$\vec\sigma$ and $\hat r$ are vector operators (irreducible tensor
operators of rank 1) use can use a general theorem for the matrix
element of a scalar product of commuting tensor operators bet ween
eigenstates of angular momentum.  Using the notation and conventions of
Edmonds \cite{Edmonds} we have
\begin{equation}
\label{matrix_element_sigma_r_1}
\fl
\Phi_{j'\ell' m'}^\dagger\,
\vec\sigma\cdot\hat r\,\Phi_{j\ell m}=(-1)^{\ell+\half+j}\delta_{jj'}
\delta_{mm'}\left\{\matrix{j&\half&\ell'\cr
                           1&\ell &\half\cr}
\right\}\,\langle{\ts\half}\|\vec\sigma\|{\ts\half}\rangle\,
\langle\ell'\|\hat r\|\ell\rangle \ .
\end{equation}
Using the conventions of Edmonds, the reduced matrix elements can be
evaluated, such that \eref{matrix_element_sigma_r_1} is
\begin{equation}
\label{matrix_element_sigma_r_2}
\fl
\Phi_{j'\ell' m'}^\dagger\,
\vec\sigma\cdot\hat r\,\Phi_{j\ell m}=(-1)^{\ell+\half+j}\delta_{jj'}
\delta_{mm'}\sqrt{6(2\ell+1)}\,
\clg10\ell 0{\ell'}0\left\{\matrix{j&\half&\ell'\cr
                                   1&\ell &\half\cr}
\right\}\ ,
\end{equation}
where we used the fact that $\clg10\ell 0{\ell'}0$ is non-zero only for
$\ell'=\ell\pm 1$.  The 6-$j$ symbol is different from zero only for
$j=\ell'\pm
\half$.  Since we have also $j=\ell\pm \half$, we have two possibilities
for a fixed $\ell$:
\begin{eqnarray}
\label{case1}
\fl
1)\quad & j=\ell+\half\quad\Longrightarrow\quad\ell'=\ell+1\ ;
\ j=\ell'-\half\\
\label{case2}
\fl
2)\quad & j=\ell-\half\quad\Longrightarrow\quad\ell'=\ell-1\ ;
\ j=\ell'+\half
\end{eqnarray}

Inserting the values of the Clebsch-Gordan coefficient $\clg10\ell
0{\ell'}0$ and of the 6-$j$ symbol into
\eref{matrix_element_sigma_r_2} we get, for both cases,
\begin{equation}
\Phi_{j'\ell' m'}^\dagger\,
\vec\sigma\cdot\hat r\,\Phi_{j\ell m}=-\delta_{jj'}\delta_{mm'}\ ,
\end{equation}
where $\ell$ and $\ell'$ are related by \eref{case1} and \eref{case2}.
Since the spinors $\Phi_{j\ell m}$ form a complete orthonormal set this
equation implies \eref{sigma_r_phi_jlm}.

\section{}

In this Appendix we obtain the spinor which is the solution of the Dirac
equation with the infinite spherical potential \eref{def_m(r)}.  To
compute the radial functions $G_{j\ell}$ and $F_{j\ell'}$ inside the
well, we first look at equations \eref{eq_G(r)_2} and \eref{eq_F(r)_2}
and make the change of variable $x={ \sqrt{E^2-m^2c^4}/(\hbar c)}\,r$.
In this way, we get equations of the form
\begin{equation}
\label{eq_Bessel}
{{\rm d}^2 f_l\over {\rm d}x^2}+{2\over x}\,{{\rm d} f_{l}\over
{\rm d}x}+\bigg(1-{l(l+1)\over x^2}\bigg)f_{l}=0\ ,
\end{equation}
where $l$ and $f_l$ stand for $\ell,\ \ell'$ and $G_{j\ell}, \
F_{j\ell'}$, respectively.  The solutions of equation \eref{eq_Bessel}
which are regular at the origin are the spherical Bessel functions of
the first kind, $j_l(x)$ (see, for instance, Abramowitz and Stegun
\cite{Abramowitz_Stegun}). Since these solutions are determined up to
an arbitrary multiplicative constant, the radial functions are
\begin{eqnarray}
G_{j\ell}=A\,j_\ell(x)\\
F_{j\ell'}=B\,j_{\ell'}(x)\ ,
\end{eqnarray}
where $A$ are $B$ are constants. We can use one of the equations
\eref{eq_G(r)} or \eref{eq_F(r)}
and the recurrence relations of the functions $j_l(x)$ (see
\cite{Abramowitz_Stegun}) to find the following relation:
\begin{equation}
B=A\,{\kappa\over|\kappa|}\sqrt{E-mc^2\over E+mc^2}\ .
\end{equation}
The complete spinor $\Psi$ then reads
\begin{equation}
\fl
\Psi_{j\kappa m}(r,\theta,\phi,t)=A\,e^{-\rmi\,E/\hbar\,t}
\pmatrix{\rmi\,j_\ell({\sqrt{E^2-m^2c^4}\over\hbar c}\,r)
\Phi_{j\ell m}(\theta,\phi)  \cr
-{\kappa\over|\kappa|}\sqrt{E-mc^2\over E+mc^2}\,\,j_{\ell'}
({\sqrt{E^2-m^2c^4}\over\hbar c}\,r)\Phi_{j\ell' m}(\theta,\phi) \cr}\ .
\end{equation}
Applying the boundary condition \eref{bound_cond} to this spinor leads
to
\begin{eqnarray}
\pmatrix{-\rmi\,{\kappa\over|\kappa|}\sqrt{E-mc^2\over E+mc^2}\,\,
j_{\ell'}(X)\Phi_{j\ell m}\cr
j_{\ell}(X)\Phi_{j\ell' m}\cr}=
\pmatrix{
\rmi\,j_\ell(X)\Phi_{j\ell m}\cr
-{\kappa\over|\kappa|}\sqrt{E-mc^2\over E+mc^2}\,\,j_{\ell'}(X)
\Phi_{j\ell' m}\cr}\ ,
\end{eqnarray}
where $X={\sqrt{E^2-m^2c^4}/(\hbar c)}\,R$ and the relation
\eref{sigma_r_phi_jlm} and its inverse were used. This equality implies
\begin{equation}
j_\ell(X)=-{\kappa\over|\kappa|}\sqrt{E-mc^2\over E+mc^2}\,\,
j_{\ell'}(X)
\end{equation}

\end{document}